\documentclass[12pt,preprint]{aastex}
\newcommand\simgt{\lower.5ex\hbox{$\; \buildrel > \over \sim \;$}}
\newcommand\simlt{\lower.5ex\hbox{$\; \buildrel < \over \sim \;$}}

\newcommand{\kmpers}{\mathrm{km\ s^{-1}}}

\newcommand{\Fig}[1]{Figure \ref{#1}}

\newcommand{\vct}[1]{\mbox{\boldmath{$#1$}}}
\newcommand{\mach}{\mathcal{M}}
\newcommand{\step}{\mathcal{H}}
\newcommand{\D}{\mathcal{D}}
\newcommand{\I}{\mathcal{I}}
\newcommand{\ext}{\mathrm{ext}}
\newcommand{\rmin}{r_\mathrm{min}}
\newcommand{\rmax}{r_\mathrm{max}}
\newcommand{\cs}{c_\mathrm{s}}
\newcommand{\tR}{\tilde{R}}

\newcommand{\tz}{\tilde{z}}
\newcommand{\sMsml}{(\mach^2-1)^{1/2}}
\newcommand{\sMsRsml}{(\mach^2\tR^2-1)^{1/2}}

\shorttitle{DYNAMICAL FRICTION OF CIRCULAR-ORBIT PERTURBER}
\shortauthors{KIM \& KIM}

\begin{document}
\title{Dynamical Friction of a Circular-Orbit Perturber in a Gaseous Medium}
\author{Hyosun Kim and Woong-Tae Kim }
\affil{Department of Physics and Astronomy, FPRD, Seoul National University, Seoul 151-742, Korea}\email{hkim@astro.snu.ac.kr,  wkim@astro.snu.ac.kr}

\begin{abstract}
We investigate the gravitational wake due to, and dynamical friction on, a 
perturber moving on a circular orbit in a uniform gaseous medium
using a semi-analytic method.  This work is a straightforward extension
of \citet{ost99} who studied the case of a straight-line trajectory.  
The circular orbit causes the bending of the wake in the background 
medium along the orbit, forming a long trailing tail.  
The wake distribution is thus asymmetric, giving rise to the drag
forces in both opposite (azimuthal) and lateral (radial) directions 
to the motion of the perturber, although the latter does not contribute to 
orbital decay much.  
For subsonic motion, the density wake with a weak tail is simply a curved 
version of that in Ostriker and does not exhibit the front-back symmetry.
The resulting drag force in the opposite direction is
remarkably similar to the finite-time, linear-trajectory counterpart.
On the other hand, a supersonic perturber is able to overtake its
own wake, possibly multiple times, and develops a very pronounced tail.
The supersonic tail surrounds the perturber in a trailing spiral fashion,
enhancing the perturbed density 
at the back as well as far front of the perturber.
We provide the fitting formulae for the drag forces as functions of the
Mach number, whose azimuthal part is surprisingly in good agreement 
with the Ostriker's formula,
provided $V_p t=2R_p$, where $V_p$ and $R_p$ are the velocity 
and orbital radius of the perturber, respectively.
\end{abstract}

\section{INTRODUCTION}

Dynamical friction refers to momentum loss suffered by a massive object 
moving through a background medium due to its gravitational interaction with 
its own induced wake.  The gravitational drag force removes angular 
momentum from an object in orbital motion, causing it to gradually 
spiral in toward the center of the orbit.  
In a pioneering study, \citet{chandra} derived the classical 
formula of dynamical friction in a uniform collisionless background,
the result of which has been applied to a number of astronomical systems.  
Examples include orbital decay of satellite galaxies orbiting their host 
galaxies
(e.g., \citealt{tremaine,lin83,weinberg, hashimoto, fujii}; see also
\citealt{bin87}),
dynamical fates of globular clusters near the Galactic center 
(e.g., \citealt{kimss03, mcm03, kimss}),
galaxy formation within the framework of hierarchical clustering scenario
(e.g., \citealt{zen03,bul05} and references therein),
formation of Kuiper-belt binaries \citep{gol02}, and 
planet migrations  \citep{planet} via interactions with planetesimals, etc.

Although less well recognized, dynamical friction also operates in gaseous 
backgrounds.  Using a time-dependent linear perturbation theory, 
\citet[hereafter O99]{ost99} derived the analytic expressions for 
the density wake and drag force 
for a perturber in a uniform gaseous medium.  O99 showed that 
resonant interactions between a perturber and pressure waves
make the gaseous drag more efficient than the collisionless drag
when the Mach number $\mach\sim1$.
She also found that even a subsonic perturber experiences a nonvanishing 
gaseous drag if interaction time between the perturber and the background
is finite.  This is an improvement on the previous notion that 
the gaseous drag is absent for subsonic perturbers 
because of the front-back symmetry in the steady-state density wake 
\citep{dokuchaev, ruderman, rep80}.

The results of O99 were confirmed numerically by \citet{sanchez99}
and have been applied to various situations including
massive black hole mergers in galactic nuclei
(e.g., \citealt{escala,liu,escala05,dotti}),  
orbital decay of compact objects (e.g., \citealt{narayan,kar01}) 
and associated viscous heating (e.g., \citealt{chang,cha03}) 
in accretion disks, and heating of an intracluster medium by 
supersonically moving galaxies in clusters 
(e.g., \citealt{elz04,fal05,kim}).  Without involving shocks,
density wakes in a collisionless medium are distributed more smoothly  
and achieve larger amplitudes than those in a gaseous medium \citep{mul83},
which led \citet{fur02} to suggest that X-ray emissions from galaxy wakes 
can in principle be used to discern the collisional character 
of dark matter in galaxy clusters.

While the results of \citet{chandra} and O99 are simple and 
provide good physical insights, they apply strictly to a mass traveling on 
a straight-line trajectory through an infinite homogeneous background.
Real astronomical systems obviously have nonuniform density distributions
and perturbers tend to follow curvilinear orbits. 
For instance, motions of galaxies in galaxy clusters, binary black holes
near the central parts of galaxies, and compact stars in accretion disks 
are better approximated by near-circular than straight-line orbits,
and their background media usually in hydrostatic equilibrium are 
stratified in the radial direction.  
Even for objects experiencing orbital decay, a near-circular orbit is a 
good approximation if the associated friction time is longer than the
orbital time.
Consideration of a circular-orbit perturber is of particular interest 
since it will allow the perturber, if supersonic, to overtake the backside 
of its wake that was created about an orbital period earlier.
In this case, a steady-state wake that eventually forms has
morphology and drag force that might be 
significantly different from 
the linear-trajectory counterparts.
Numerical simulations carried out by \citet{sanchez} and \citet{escala} 
indeed show that the density wake by a near-circular orbit perturber
contains a trailing spiral tail, which is absent in the linear-trajectory 
cases.  
They also showed that the resulting drag force is smaller than the 
estimate based on the formula given in O99,
which is probably due to the near-circular orbit,
although the effect of nonuniform backgrounds in their models 
cannot be ignored completely.

The drag formula based on perturbers moving straight in either a
collisionless medium or a gaseous medium depends on the Coulomb logarithm 
$\ln(\Lambda) \equiv \ln(\rmax/\rmin)$, where
$\rmin$ and $\rmax$ are the cutoff radii introduced 
to avoid a divergence of the force integrals.\footnote{For a perturber
moving with velocity $V_p$ through a gaseous medium, $\rmax=V_p t$, 
where $t$ denotes time elapsed since the perturber was introduced (O99).}
While many previous studies conventionally adopted $\rmin$ and $\rmax$ as 
the characteristic sizes of the perturber and the background medium,
respectively, the choice of $\rmax$ remains somewhat
ambiguous for objects moving on near-circular orbits
(e.g., \citealt{bin87}).
For {\it collisionless} backgrounds, \citet{hashimoto} and \citet{fujii} 
performed N-body experiments for orbital evolution of satellite galaxies
in a spherical halo, and found that 
the Coulomb logarithm with $\rmax$ varying proportionally
to the orbital radius rather than fixed to the system size
gives better fits to their numerical results (see also \citealt{tremaine}).
We shall show in the present work that 
a similar modification of the Coulomb logarithm is necessary in order 
to apply the results of O99 to perturbers on near-circular orbits 
in {\it gaseous} backgrounds, as well.

In this paper, we consider a perturber moving on a circular orbit in a 
uniform gaseous medium.  Using a linear semi-analytic approach, we 
explore the structure of the density wake,
evaluate the drag force on the perturber, 
and compare them with those in the straight-line trajectory cases.  
In \S2, we revisit the linear perturbation
analysis of O99 for the perturbed density response and apply it to
the case of a circular-orbit perturber. We solve the resulting
equations numerically.  In \S3, we present the numerical results 
for the wakes and drag forces with varying Mach number.
We provide simple fitting expressions to the numerical results, 
and show that Ostriker's formula still gives a good estimate for 
the drag force on a circular-orbit perturber only if 
the outer cutoff radius in the Coulomb logarithm is taken equal to the 
orbital diameter of the perturber.
In \S4, we summarize the present work and briefly discuss our finding. 

\section{FORMULATION}

\subsection{Formal Solution for Density Wake}

We consider the response of gas to 
a point-mass perturber moving on a circular orbit and calculate the 
resulting gravitational drag force on the perturber.  
We treat the gas as an inviscid, adiabatic fluid, and 
do not consider the effects of magnetic fields as well 
as gaseous self-gravity.
The governing equations for ideal hydrodynamics are
\begin{equation}\label{eq:con}
  \frac{\partial\rho}{\partial t}+\vct{\nabla}\cdot(\rho\vct{v})=0,
\end{equation}
and
\begin{equation}\label{eq:mom}
  \frac{\partial\vct{v}}{\partial t}+\vct{v}\cdot\vct{\nabla}\vct{v}
  = -\frac{1}{\rho}\vct{\nabla}P-\vct{\nabla}\Phi_\ext,
\end{equation}
where $\Phi_\ext$ is the gravitational potential of the perturber.
Other symbols have their usual meanings. 
  
Following O99,
we consider an initially uniform gaseous medium with density $\rho_0$.
Assuming that the wake induced by the perturber remains at small amplitudes, 
we linearize equations (\ref{eq:con}) and (\ref{eq:mom}) using
$\rho=\rho_0[1+\alpha(\vct{x},t)]$ and 
$\vct{v}=\cs\,\vct{\beta}(\vct{x},t)$, 
where $c_s$ is the adiabatic speed of sound in the unperturbed medium 
and $\alpha$ and $\vct{\beta}$ denote the 
dimensionless density and velocity perturbations, respectively. 
Eliminating $\vct{\beta}$ from the linearized equations, one 
obtains a three-dimensional wave equation
\begin{equation}\label{eq:wave}
  \vct{\nabla}^2\alpha-\frac{1}{\cs^2}\frac{\partial^2\alpha}{\partial t^2}
  = -\frac{4\pi G}{\cs^2}\rho_\ext(\vct{x},t), 
\end{equation}
where $\rho_\ext = \vct{\nabla}^2\Phi_\ext /(4 \pi G)$ represents 
the mass density of the perturber.
The formal solution to equation (\ref{eq:wave}) based on the 
the retarded Green function technique is given by
\begin{equation}\label{eq:formal}
  \alpha(\vct{x},t) = \frac{G}{\cs^2} \int\!\!\!\int d^3x' dt'
  \ \rho_\ext(\vct{x'},t')
  \frac{\delta\,[t'-(t-\vert\vct{x}-\vct{x'}\vert/\cs)]}
       {\vert\vct{x}-\vct{x'}\vert}
\end{equation}
(O99; see also \citealt{jackson}).
In the case of a perturber on a straight-line trajectory,
O99 solved equation (\ref{eq:formal}) directly to obtain an 
expression for the perturbed density.  The same result was found 
by \citet{fur02} who independently used a Fourier transform method
in both space and time variables. 

\subsection{Density Wake for Circular-Orbit Perturbers}

We now concentrate on the case where a point-mass perturber
with mass $M_p$ moves on a circular orbit with a fixed orbital radius 
$R_p$ and a constant velocity $V_p$ in an otherwise uniform gaseous medium;  
the angular speed of the perturber is $\Omega = V_p/R_p$.
It is convenient 
to work in cylindrical coordinates ($R$, $\varphi$, $z$) whose origin lies 
at the center of the orbit.  The $\hat{z}$-axis points perpendicular to the 
orbital plane.  Assuming that the perturber
is introduced at $(R_p, 0, 0)$ when $t=0$, one can write 
$\rho_\ext(\vct{x},t)=M_p\; 
\delta (R-R_p)\, \delta [R_p(\varphi- \Omega t)]\, 
\delta (z)\, \step (t)$, 
where $\step (t)$ is a Heaviside step function. 
Equation (\ref{eq:formal}) is then reduced to 
\begin{equation} \label{eq:delta}
  \alpha(\vct{x},t)=\frac{G M_p}{\cs^2 R_p} \int dw\ 
      \frac{\delta\left(w+s+\mach\ d\,(w;\tR,\tz)\right)}{d\,(w;\tR,\tz)}\ 
      \step\left(\frac{w+\varphi}{\Omega}\right),
\end{equation}
where $w \equiv \varphi^\prime-\varphi$ and 
$s \equiv \varphi-\Omega t$ are angular distances in 
the $z=0$ plane\footnote{Note that $w$ and $s$ in O99 are 
defined as {\it linear} distances along the line of motion, 
while they measure {\it angular} distances in the present work.},
\begin{equation}\label{eq:dist}
d\,(w;\tR,\tz)\equiv \frac{\vert\vct{x}-\vct{x'}\vert}{R_p} 
=\left(1 + \tR^2+\tz^2-2\tR\cos w\right)^{1/2},
\end{equation}
and $\mach\equiv V_p/\cs$ is the Mach number of the perturber.
In equation (\ref{eq:dist}), $\tR\equiv R/R_p$ and $\tz\equiv z/R_p$.

\Fig{sketch} schematically illustrates the situation at the orbital plane
and the meanings of variables used in equation (\ref{eq:delta}). 
At time $t$, the perturber is located at $\vct{x_p}=(R_p, \Omega t, 0)$.  
During its journey along the thick curve, the perturber continuously
launches sound waves that propagate into the background gaseous medium.  
The position $\vct{x}=(R,\varphi,z)$ denotes a region of interest
in the surrounding gas where the density response will be evaluated.
Since the sound waves have finite traveling time, only the signals 
emitted by the perturber at the location(s) $\vct{x'}=(R_p, \varphi', 0)$ 
at the retarded time $t'=t-\vert\vct{x}-\vct{x'}\vert/\cs$ 
are able to affect the point $\vct{x}$ at time $t$. 
Note that $s$ and $w$ represent the projected angular distances 
in the orbital plane between $\vct{x_p}$ and $\vct{x}$ and 
between $\vct{x'}$ and $\vct{x}$, respectively.  The symbol $d$ 
in equation (\ref{eq:dist}) refers to the three-dimensional linear distance
between $\vct{x'}$ and $\vct{x}$ normalized by $R_p$.

Using the identity $\delta\,[f(w)]=\sum_i\delta(w-w_i)/\vert f'(w_i)\vert$, 
where $w_i$ are the roots of an arbitrary function $f(w)$, equation 
(\ref{eq:delta}) is further simplified to
\begin{equation}\label{eq:alpha}
 \alpha( \vct{x},t) 
  = \frac{G M_p}{\cs^2 R_p}\ \D (\vct{x},t),
\end{equation}
with the dimensionless perturbed density 
\begin{equation} \label{eq:D}
  \D(\vct{x},t) = \sum_{w_i}
  \frac{\mach} {\vert w_i+s-\mach^2\tR\ \sin w_i \vert}
  \step\left(\frac{w_i+\varphi} {\Omega}\right).
\end{equation}
Here, the summation is over all possible roots $w_i$ that satisfy 
the condition 
\begin{equation} \label{eq:w} 
\mach\ d\,(w_i;\tR,\tz) = - ( w_i+s ),
\end{equation}
for fixed values of $\tR$, $\tz$, and $s$.  
The function $\step [(w_i+\varphi)/\Omega]$ in equation (\ref{eq:D}) 
defines the region of influence (or casual region) outside of which sonic 
perturbations sent off by the perturber at $t=0$ have insufficient time 
to reach.  For $\tR, \tz, \Omega t\gg 1$, equation (\ref{eq:w}) yields 
$w_i + \varphi = \Omega t - \mach (\tR^2+\tz^2)^{1/2}$, so that 
the region of influence corresponds roughly to a sphere with radius $c_st$
centered at the orbital center.
Appendix \ref{sec:limit} presents limiting solutions of 
equation (\ref{eq:D}) near the perturber. 

Since $d$ is a periodic function of $w$ with period 2$\pi$, 
equation (\ref{eq:w}) has at least one real root and may possess multiple
roots for $w_i$ depending on the values of $\mach$, $\tR$, $\tz$, and $s$.
In Appendix \ref{sec:wi}, we describe how the number of solutions of
equation (\ref{eq:w}) vary with the Mach number of a perturber.
It turns out that there is only a single root for $w_i$ 
everywhere for a subsonic, circular-orbit perturber, which is the same as 
in the straight-line trajectory case (O99). 
When a circular-orbit perturber moves at a supersonic speed, however, 
equation (\ref{eq:w}) has an odd number of roots that contribute to
the wake in a steady state, which is distinct from the straight-line 
trajectory case where
only one or two points along the orbit influence the wake. 

\subsection{Gravitational Drag Force}

Once the gravitational wake $\alpha( \vct{x},t)$ is found, 
it is straightforward to evaluate the drag force 
exerted on the perturber: 
\begin{equation}
  \vct{F}_\mathrm{DF} = G M_p \rho_0 \int d^3\vct{x}\ 
  \frac{\alpha(\vct{x},t)\ (\vct{x}-\vct{x_p})} 
       {\vert \vct{x}-\vct{x_p} \vert^3}\mbox{.}
\end{equation}
In the straight-line trajectory case studied by O99,
$\alpha(\vct{x},t)$ always remains cylindrically symmetric with respect to
the line of motion, resulting in the drag force 
in the anti-parallel direction. 
When the perturber is on a circular orbit, $\alpha(\vct{x},t)$ 
loses the cylindrical symmetry and instead becomes symmetric relative
to the orbital plane, making the vertical component of 
$\vct{F}_\mathrm{DF}$ vanish.  

We decompose the nonvanishing parts into
the radial and azimuthal components:  
\begin{equation}\label{eq:fdf}
\vct{F}_\mathrm{DF} =  -\mathcal{F} \;
(\I_{\rm R} \hat{\vct{R}} + \I_{\rm \varphi} \hat{\vct{\varphi}}),
\;\;\;\;
\mathcal{F} \equiv  \frac{4\pi\rho_0 (GM_p)^2}{V_p^2} 
\end{equation}
where 
\begin{mathletters} \label{eq:I}
  \begin{equation}
    \I_{R} = -\frac{\mach^2}{4\pi} \int d^3\tilde{\vct{x}}\ 
    \frac{\D({\vct{x}},t)\ (\tR\cos s -1)}
	 {(1+\tz^2+\tR^2-2\tR\cos s)^{3/2}}\mbox{,}
  \end{equation}
and
  \begin{equation}
    \I_{\varphi} = -\frac{\mach^2}{4\pi} \int d^3\tilde{\vct{x}}\ 
    \frac{\D({\vct{x}},t)\ \tR\sin s}
	 {(1+\tz^2+\tR^2-2\tR\cos s)^{3/2}}.
  \end{equation}
\end{mathletters}
Note that 
$\I_{R}$ measures the drag force along the lateral direction of the 
instantaneous perturber motion, 
while $\I_{\varphi}$ is for the backward direction.  
The dimensional term $\mathcal{F}$ in equation (\ref{eq:fdf}) allows to
directly compare $\I_{R}$ and $\I_{\varphi}$ with the
linear-trajectory counterparts (see eq.~[12] of O99).
As we shall show in \S\ref{sec:sum}, 
it is the azimuthal drag $\I_{\varphi}$ that 
is responsible for the orbital decay of a perturber.

\subsection{Numerical Method}

We solve equations (\ref{eq:D}) and  (\ref{eq:w}) 
numerically
to find the perturbed density distribution $\D(\vct{x},t)$ for given
$\mach$ and $t$. 
We first construct a three-dimensional Cartesian mesh centered at the
center of the orbit, and solve equation (\ref{eq:w}) for
$w_i$ at each grid point using a hybrid Newton-bisection method.  
By checking the conditions for multiple roots discussed 
in Appendix \ref{sec:wi}, 
we ensure that we do not miss any solution for $w_i$.
The corresponding drag forces $\I_R$ and $\I_\varphi$ are calculated
by direct integration of equations (\ref{eq:I}).

Since the density wake often exhibits sharp discontinuities especially
for supersonic perturbers and is distributed over a large spatial range,
it is important to check that the drag forces we calculate do not 
depend on the size of the computational box and its resolution. 
For fixed $\mach$, we repeated the calculations with varying box size
and resolution and found that depending on the Mach number, the box size 
of $\sim (20-100)R_p$ and resolution of $\sim (80-640)$ grids per 
$1R_p$ are sufficient to guarantee good convergence of the drag forces.
Although the density perturbations are non-zero outside the box,
they have very low amplitudes and are located far from the perturber, 
providing a negligible contribution to the drag.
Very high resolution calculations are required for Mach numbers 
near the critical values $\mach_n$, in which cases the wake tails become
thin and dense (see Appendix \ref{sec:wi}).

\section{RESULTS}\label{sec:res}

\subsection{Density Wake} \label{sec:density}

\subsubsection{Supersonic Cases}\label{sec:superwake}

We begin by illustrating temporal evolution of density perturbations 
induced by a supersonic perturber.
\Fig{evolution} shows snapshots of the density wake and the corresponding 
number of roots of equation (\ref{eq:w}) at the orbital 
plane for $\mach=2.0$. 
Time is expressed in units of $R_p$/$c_s$. The black circle represents 
the orbit of a perturber which, introduced at 
$ (R, \varphi, z) =(R_p, 0, 0)$ initially,
moves in the counterclockwise direction.
At early time ($t\simlt 1.5$), the density wake consists of a Mach cone
and a sonic sphere that are curved along the orbit.
Except the bending, the overall wake structure, not to mention the number
of solutions for $w_i$ which is either one or two inside the casual 
region, is the same as in the straight-line trajectory case of O99.
As the wake bends further, the Mach cone and the sonic sphere become 
folded at the innermost interface, creating high-density regions near 
the center ($t=1.8$).  The wake expands with time 
at a sonic speed and the Mach cone becomes elongated further. 

Unlike in the case of a straight-line trajectory 
where the Mach cone and the sonic sphere never interact with each other, 
the perturber (and the head of the Mach cone) on a circular orbit is 
able to enter its own wake, 
providing additional perturbations for some regions inside the sonic sphere.
Alternatively, this can be viewed as the sonic sphere whose center lies 
at the initial position of the perturber expands radially outward, 
swallowing a part of the elongated Mach cone.  Consequently, the overlapping 
of the Mach cone and the sonic sphere creates a high-density trailing tail
that has received perturbations three times from the perturber
($t\simgt 1.9$).
\Fig{evolution} shows that immediate outside the sonic sphere, 
there still exists a region of the undisturbed Mach cone 
with two $w_i$'s.  As time proceeds, however, this region
moves away from the perturber and thus gives an increasingly small 
contribution to the drag force.  In a steady state which is 
attained at $t\rightarrow\infty$, the entire domain is affected  
either once or three times by the perturber.

\Fig{steadyM20} displays the steady-state distributions of the density wake 
for $\mach=2.0$ on the $x/R_p=1$, $y=0$, and $z=0$ planes, 
which clearly shows that the trailing tail loosely wraps around 
the perturber.
The tail becomes narrower with increasing $|z|$.
A close inspection of the tail at the $z=0$ plane shows that density 
becomes smaller away from the perturber along the tail and is largest
at the edges across the tail.  This is similar to the density
distribution inside the Mach cone in the linear trajectory
case where diverging flows (and reduced gravitational potential) after 
the shock cause density to decrease away from the shock front
(and the perturber). 
This suggests that the edges of the tail are shock fronts.
The outer edge of the tail shown in \Fig{steadyM20} that connects smoothly
all the way to the perturber corresponds indeed to the surface of the 
curved Mach cone (see also Fig.~\ref{sketch}).  
On the other hand, the inner edge of the tail traces the interface between 
the Mach cone and the sonic sphere that is newly swept up by the expanding
sonic sphere (see Fig.~\ref{evolution}).
Appendix \ref{sec:angle} proves that the half-opening angle of the
head of the curved Mach cone in the $z=0$ plane is equal to 
$\sin^{-1} (1/\mach)$, entirely consistent with the case of a linear 
trajectory.

The wake tail becomes thicker as $\mach$ increases from unity. 
\Fig{overlap} shows the steady-state density wakes 
as well as the number of $w_i$'s that contribute to $\D$
for $\mach=4.0$ and $\mach=5.0$ at the $z=0$ plane.  In both panels, 
the perturber moving in the counterclockwise direction is located 
at $x/R_p=1, y=0.$
When $\mach=4.0$, the tail is fat enough to cover most of the space
except near the center and a narrow lane between the tail edges.  
At $\mach_1 \approx 4.603$, the inner edge of the tail
eventually touches the neighboring outer edge, 
enabling three $w_i$ for the entire region under the conditions 
expressed by equation (\ref{eq:ncond}).
When the Mach number is slightly larger than $\mach_1$,
the tail overlaps itself.  This in turn creates a new tail with 
five $w_i$'s, as \Fig{overlap}\textit{b} displays.  
As $\mach$ increases further, 
the new tail again becomes thicker, starts to overlap
at $\mach_2 \approx 7.790$, and produces a narrow lane with
seven $w_i$'s when $\mach>\mach_2$.  The same pattern repeats with 
increasing $\mach$, and equation (\ref{eq:M}) determines 
the critical Mach numbers.

\subsubsection{Subsonic Cases}\label{sec:subwake}

Unlike in the supersonic cases where a perturber can overtake its own wake 
and create a tail with complicated structure, sonic perturbations generated 
by a subsonic perturber produce a gravitational wake that is spatially smooth
and does not involve a shock.  Since perturbations propagate 
faster than a perturber with $\mach<1$, 
the whole casual region is affected by a perturber 
just once, corresponding to a single $w_i$ at any position.
\Fig{steadyM05} shows the slices
of the perturbed density for $\mach=0.5$ in the
$x/R_p=1$, $y=0$, and $z=0$ planes when a steady state is reached.
Again, the perturber is located at $x/R_p=1$, $y=z=0$.
The density structure in the $z=0$ plane is simply a curved version 
of what a linear-trajectory perturber would produce. 
In particular, as Appendix \ref{sec:limit} shows, 
the iso-density surfaces near the perturber have the shapes of 
oblate spheroids with ellipticity $e=\mach$, with the short axis parallel to
the direction of the motion of the perturber, a characteristic feature
of a subsonic wake created by a linear-trajectory perturber (O99).

Notice, however, that the bending of wakes in circular-orbit cases 
makes the perturbed density distributions intrinsically asymmetric.
This results in nonvanishing drag forces even in a steady state,
and the dominant contribution to the drag comes from high-density
regions near the perturber.
This is markedly different from the purely steady-state linear-trajectory 
cases where a subsonic perturber experiences no drag due to
the front-back symmetry of a wake about the perturber 
(e.g., \citealt{rep80}).
Even if the finite interaction time between the straight-line perturber and
the background gas is considered, regions with symmetric perturbed
density close to the perturber exert zero net force (O99).
Nevertheless, the resulting drag force in the backward direction of motion 
on a circular-orbit perturber is almost the same as that in the 
linear-trajectory cases, as we will show in the next subsection. 
Compared with supersonic cases, the tail in a subsonic wake is short,
loosely wound, and very weak, suggesting that its contribution to
the drag force is negligible.

\subsection{Gravitational Drag Force}

As sonic perturbations launched from a perturber at $t=0$ propagate 
radially outward, the volume of space exerting the gravitational 
drag on the perturber grows with time.
\Fig{time} plots the drag force as functions of time for a few chosen
Mach numbers.  The solid and dotted lines are for $\I_\varphi$ and
$\I_R$, respectively.  Since $\D(\vct{x},t)$ is singular at 
$\vct{x}=\vct{x}_p$,  only the region with $r > r_{\rm min} = R_p/10$ 
is taken into account in force computation,
where $r$ is the three-dimensional distance measured from the perturber;
the dependence on $\rmin$ will be checked below.
The drag force on a subsonic perturber increases almost linearly with
time before turning abruptly to a constant value, whereas a supersonic
wake with a high-density tail gives rise to slow fluctuations 
in the drag at early time.  At any event, both components of the drag 
force converge to respective steady-state values, 
typically within the sound crossing time
over the distance equal to the orbital diameter or 
within about an orbital period when $\mach$ is of order unity.
The primary reason for this is of course because the perturbed density 
decreases quite rapidly with $r$ and also because
gravity is an inverse-square force.
This is unlike the case of a straight-line trajectory where
the drag increases secularly as $\ln(V_p t)$ for $\mach>1$.
The fast convergence of the drag force guarantees that 
one can use the steady-state values of $\I_\varphi$ and $\I_R$ 
for all practical purposes.

Next, we check the dependence of the drag force on $\rmin (\ll R_p)$.
\Fig{rmin} shows the results for 
$\mach=0.5$, 2.0, and 4.0.  The sizes of errorbars associated with 
finite grid resolution are smaller than the radius of a solid circle
at each data point.
First of all, the drag force, $\I_R$, in the radial direction converges 
to a constant value as $\rmin$ decreases for both subsonic and 
supersonic cases. 
On the other hand, the drag force, $\I_\varphi$, in the opposite direction
of the orbital motion, varies as $\ln (1/\rmin) $ with decreasing $\rmin$ for
supersonic perturbers, while independent of $\rmin$ for subsonic
perturbers.
This dependence of $\I_\varphi$ on small $\rmin$ for 
circular-orbit perturbers is exactly the same as in the linear 
trajectory cases, which makes sense because the curvature of
a circular orbit is almost negligible in a tiny volume 
near the perturber.

We plot in \Fig{forces} the steady-state drag forces for a 
circular-orbit perturber as functions of the Mach number.  
For all the points, $\rmin=R_p/10$ is taken and numerical 
convergence is checked.  
Filled circles give $\I_\varphi/\mach^2$, while open circles are for 
$\I_R/\mach^2$, which can be compared with Figure 3 of O99.
For practical use, we fit the data using 
\begin{equation}\label{eq:IR}
  \I_R = \left\{\begin{array}{l l@{\ }r@{\;}c@{\,}l}
    \mach^2\ 10^{\ 3.51\mach-4.22},   &\textrm{for}&&\mach&<1.1,\\
    0.5\  \ln\big[9.33\mach^2(\mach^2-0.95)\big],
                                       &\textrm{for}&1.1\leq&\mach&<4.4,\\
    0.3\ \mach^2,                       &\textrm{for}&4.4\leq&\mach,
  \end{array}\right.
\end{equation}
and
\begin{equation}\label{eq:Iphi}
  \I_\varphi = \left\{\begin{array}{l l@{\ }r@{\;}c@{\,}l}
     0.7706\ln\left(\frac{1+\mach}{1.0004-0.9185\mach}\right)
       -1.4703\mach,
                     &\textrm{for}&&\mach&<1.0,\\
     \ln\left[330 (R_p/\rmin) (\mach-0.71)^{5.72}\mach^{-9.58} \right],
                      &\textrm{for}&1.0\leq&\mach&<4.4, \\
     \ln\left[(R_p/\rmin)/(0.11\mach+1.65)\right],
                     &\textrm{for}&4.4\leq&\mach,
  \end{array}\right.
\end{equation}
which are drawn as solid lines in \Fig{forces}.
The fits are within 4\% of our semi-analytic results for $\mach<4.4$ 
and within 16\% for $\mach>4.4$. 
Both components of the drag force peak at $\mach\sim 1.2-1.4$, 
analogous to the linear-trajectory cases, and $\I_\varphi$ dominates 
over $\I_R$ for transonic perturbers.  
Although $\I_R$ is larger than $\I_\varphi$ 
for $\mach \simgt 2.2$, its effect on orbital decay of a perturber 
is insignificant (see \S\ref{sec:sum}).
The local bumps in the drag forces at $\mach \approx 4.6$ and 7.8 are due to
the self-overlapping of a wake tail explained in \S\ref{sec:superwake}.

\Fig{forces} also plots as dotted curves the results of O99 for drag force
on linear-trajectory perturbers. 
Despite the difference in the shape of the orbits,
the agreement between $\I_\varphi$ and 
Ostriker's formula is excellent for the subsonic case.
This presumably reflects the fact that other than bending, the wake 
structure created by a subsonic circular-orbit perturber is not 
significantly different from the linear-trajectory counterpart 
(see \S\ref{sec:subwake}). 
Note also that Ostriker's formula for the supersonic case,
for which $V_pt=2R_p$ is adopted to fit our numerical results, 
is overall in good agreement with $\I_\varphi$ for a range of 
$\mach$. 
It is remarkable that the subsonic and supersonic drag formulae 
(with $V_p t$ chosen adequately in supersonic cases) 
obtained from perturbers moving straight still give a reasonably good 
estimate for the drag even on circular-orbit perturbers.

\section{SUMMARY AND DISCUSSION}\label{sec:sum}

We have taken a semi-analytic approach to study the gravitational wake
and the associated drag force on a perturber moving on a circular orbit 
in an infinite, uniform gaseous medium.  This work extends 
\citet{ost99} who studied the cases with straight-line
trajectory perturbers.  The circular orbit generally causes the wake to 
bend along the orbit and creates a trailing tail.
For a subsonic perturber, the density wake has a weak tail and is 
distributed quite smoothly (see Fig.~\ref{steadyM05}). 
On the other hand, a supersonic perturber can catch up with its own wake,
possibly multiple times depending on the Mach number, 
forming a very pronounced trailing tail across which density 
changes abruptly (see Fig.~\ref{steadyM20}).  
Although the region influenced by the perturber keeps expanding 
with time into the surrounding medium, the drag
force promptly saturates to a steady state value within less than 
the crossing time of sound waves over the distance equal to 
the orbital diameter.

Because of asymmetry in the density wake, it is desirable to decompose 
the drag force into two components: $\I_R$ and $\I_\varphi$ in the radial and
azimuthal directions, respectively (eq.~[\ref{eq:fdf}]).
\Fig{forces} plots our main results for 
$\I_R$ and $\I_\varphi$ as functions of the Mach number $\mach$;
equations (\ref{eq:IR}) and
(\ref{eq:Iphi}) give the algebraic fits to the numerical results. 
The azimuthal drag force varying rather steeply with $\mach$ peaks 
at $\mach\sim1.2-1.4$, while the radial force is highly 
suppressed at $\mach<1$ and exceeds the azimuthal drag at 
$\mach\simgt 2.2$.  
It is remarkable that the drag on linear-trajectory perturbers given in 
O99 is almost identical to 
$\I_\varphi$ for subsonic cases, and gives a good approximation
for supersonic cases, too, provided $V_pt = 2R_p$.

A striking feature in gravitational wakes formed by circular-orbit
perturbers is a long tail in a trailing spiral shape.  Such a tail 
structure is indeed commonly seen in recent hydrodynamic simulations 
for black hole mergers in a gaseous medium 
(e.g., \citealt{escala,escala05,dotti}; see also \citealt{sanchez}).
Albeit much weaker, it is also apparent in N-body simulations 
for satellite orbital decay
in a collisionless background (e.g., \citealt{weinberg,her89}).
For supersonic perturbers, the tail is a curved Mach cone and 
bounded by the shock fronts that propagate radially outward.
As explained in Appendices \ref{sec:wi} and \ref{sec:angle}, the 
outer edge of the tail is described by $\Omega t - \varphi = w_+ + y_+$,
where $w_+$ and $y_+$ are defined through equation (\ref{eq:wy}).  
By taking a time derivative of this equation for fixed $\varphi$,
one can show that the propagation speed $\dot R_{\rm sh}$ 
of the spiral tail in the radial direction is given by
$\dot R_{\rm sh}/c_s = \mach \tR(\mach^2\tR^2-1)^{-1/2}$
for $\tR\equiv R/R_p\geq1$.  
Clearly, $\dot R_{\rm sh}=c_s$ for $R/R_p \gg 1$,
and $\dot R_{\rm sh} = c_s\mach/(\mach^2-1)^{1/2}$ near $R/R_p =1 $ as 
equation (\ref{eq:half}) implies.
This prediction is roughly consistent with 
\citet{escala} who empirically found that the tail in a model 
with $\mach=\sqrt{2}$ has an average propagation speed of 
$\dot R_{\rm sh} \approx 1.1 c_s$.

Many previous studies on dynamical friction 
adopted the drag formula based on perturbers moving straight and 
estimated the Coulomb logarithm by taking $\rmax=R_{\rm sys}$, 
where $R_{\rm sys}$ is the system size (see references in \S1). 
Unlike a straight-line trajectory,
a circular orbit naturally introduces a characteristic 
length scale, $R_p$. The results of our semi-analytic analyses suggest 
that, at least in a gaseous medium, 
the drag force obtained for linear-trajectory perturbers 
can be a reasonable approximation to that for circular-orbit 
perturbers with the same Mach number 
if $\rmax \equiv V_p t$ is taken equal to $2R_p$.
Since $R_{\rm sys}>2R_p$ in most relevant situations, using 
$\rmax =R_{\rm sys}$ usually overestimates the drag force for 
objects in near-circular motion. 
From hydrodynamic simulations for orbital decay of 
perturbers in a stratified gaseous sphere, 
\citet{sanchez} suggested $\ln(\Lambda)=
\ln(\rmax/\rmin) \rightarrow 2.34 \ln 
(0.79 R_p/\rmin)$ with an identification $\rmin = 2.25 R_{\rm soft}$, 
where $R_{\rm soft}$ is the softening radius of the point-mass potential
\citep{sanchez99}. This happens to be similar to our suggestion 
$\ln(\Lambda) \rightarrow \ln(2R_p/\rmin)$ for the parameter
range $R_p/\rmin \sim 2-6$ that they considered.  

Our suggestion for the Coulomb logarithm on near-circular orbit perturbers 
in a gaseous medium is also consistent with the common prescription 
for orbital decay of small satellites in collisionless backgrounds
(e.g., \citealt{tremaine, lin83, hashimoto, fujii}).
In particular, \citet{hashimoto} found that 
Chandrasekhar's formula with $\ln(\Lambda) \rightarrow
\ln[R_p/(1.4 R_{\rm soft})]$ gives excellent fits to the results of
their N-body simulations.
The discrepancy between our suggestion and their prescription may be
attributable in part to the effects of 
collisionless nature, nonuniform background density, 
and self-gravity of particles in their models. 

Density inhomogeneity in the background can also make significant 
changes to the classical drag force.  For collisionless backgrounds,
\citet{just} found that nonuniform density induces 
an additional drag force in the lateral direction of 
the perturber motion, taking up to 10\% of the drag in the 
backward direction.
They proposed the inverse of a local density gradient as 
an appropriate choice for $\rmax$ in the Coulomb logarithm 
(see also \citealt{spi03}), which was confirmed by \citet{are07} who
ran a number of numerical simulations for systems with a 
large central density concentration.
Hydrodynamic models in \citet{sanchez} studied the combined effects of 
nonuniform backgrounds and curvilinear orbits 
on dynamical friction in a gaseous medium,
although it is challenging to isolate each effect separately.

Finally, as a heuristic example, we consider the dynamical friction 
of a galaxy on a near-circular orbit subject to the drag force given by
equations (\ref{eq:IR}) and (\ref{eq:Iphi}) due to an intracluster
medium.  The cluster is dominated by dark matter 
whose mass distribution follows the NFW profile with the characteristic mass
$M_0=6.6\times 10^{14} M_\sun$ and the scale radius $r_\mathrm{s}=460$ kpc
\citep{NFW}.  The intracluster medium has 
a constant electron density $n_e=0.05\ \mathrm{cm}^{-3}$ 
and temperature 1 keV; the corresponding adiabatic speed of sound is 
$\cs = 500\ \kmpers$.  
Initially, a galaxy with size $\rmin=10$ kpc and 
mass $M_p=5\times 10^{11} M_\sun$ including 
a dark halo (e.g., \citealt{zen03,halkola}) is orbiting at $R_0=100$ kpc 
with an equilibrium velocity $V_p=10^3\ \kmpers$ about the
cluster center.
The equations of motion in the orbital plane are
\begin{mathletters} \label{eq:gal}
  \begin{equation}
    \ddot{R} - R\dot{\varphi}^2 = -(\mathcal{F}/M_p) \I_R - 
    \frac{d\Phi_{\rm NFW}}{dR}
     \mbox{,}\\
  \end{equation}
  \begin{equation}
    R\ddot{\varphi} + 2\dot{R}\dot{\varphi}  = 
    -(\mathcal{F}/M_p) \I_\varphi,
   \end{equation}
\end{mathletters}
where $\Phi_{\rm NFW}$ is the NFW potential and dots represent 
derivatives with respect to time.  
\Fig{orbitNFW} plots the 
resulting temporal variations of the orbital radius of the galaxy.
The solid line corresponds to the case with full $\I_R$ and $\I_\varphi$
given by equations (\ref{eq:IR}) and (\ref{eq:Iphi}),
while the dashed line is for a controlled case 
where $\I_R$ is artificially taken to be zero.  
Except for slight mismatches in phase,
both agree fairly well with each other, demonstrating that 
the radial drag force does not have a serious impact on the orbital decay.  
As equation (\ref{eq:gal}) implies, it is the azimuthal
drag $\I_\varphi$ that takes away most of the orbital angular momentum from 
the galaxy;  the radial drag changes the orbital eccentricity more 
than the semi-major axis (\citealt{bur76}; see also \citealt{just}).

\Fig{orbitNFW} also plots as dot-dashed line the decay of the orbital radius
blindly using Ostriker's formula with $\ln (V_pt/\rmin) = 4.6$,
corresponding to $V_pt=1$ Mpc.  
While the galaxy motion remains supersonic ($t<1.5$ Gyr),
this value of the Coulomb logarithm overestimates the realistic drag force,
on average, by a factor of 1.7 and thus brings the galaxy 
to $R=0.1R_0$ in 2 Gyrs, 
which is about 2.3 times faster than the estimate based on the realistic 
drag force. On the other hand, the result of using $V_p t = 2R(t)$ in 
the formula of O99, shown as dotted line in \Fig{orbitNFW}, is in excellent 
agreement with those using equations (\ref{eq:IR}) and (\ref{eq:Iphi}).
This demonstrates again that Ostriker's formula with $V_p t = 2R$ 
gives quite a good approximation to the drag force even in the case of
a circular-orbit perturber.

\acknowledgments

We are grateful to E.\ Ostriker for suggesting this project 
and making stimulating comments.  We also acknowledge a thoughtful report
from the referee, A.\ Escala, and helpful comments from J.\ 
S{\'a}nchez-Salcedo.
This work was supported by Korea Science and Engineering
Foundation (KOSEF) grant R01-2004-000-10490-0 at SNU. 
H.\ Kim has been partially supported by the BK21 project of the 
Korean Government.
The numerical computations presented in this work were performed on the 
Linux cluster at KASI (Korea Astronomy and Space Science Institute) 
built with funding from KASI and ARCSEC.

\appendix
\section{
Solutions for $\D(\vct{x},t)$ Near the Perturber}
\label{sec:limit}

In this Appendix, we explore the approximate solutions to equation 
(\ref{eq:D}) near the perturber ($|\tR-1|, |s|, |\tz| \ll 1$). 
For subsonic motion ($\mach<1$), $|w_i| \ll 1$ and 
$d\approx ( \delta^2 + w^2 + \tz^2)^{1/2}$ with $\delta\equiv \tR-1$. 
In this case, equation (\ref{eq:w}) has a single root, 
\begin{equation}\label{eq:www}
w_1 = \frac{s + \mach [ s^2 + (1-\mach^2)(\delta^2 + \tz^2)]^{1/2}}
{\mach^2-1}.
\end{equation}
Inserting $w_1$ into equation (\ref{eq:D}) 
and retaining the 
first-order terms in  $|\tR-1|, |s|$, and $|\tz|$, one obtains
$\D(\vct{x}) 
= [s^2 + (1-\mach^2)(\delta^2 + \tz^2)]^{-1/2}$ at a steady state,
whose functional form is identical to equation (9) of O99 
for $\mach<1$.  Obviously,
the equi-density surfaces of this distribution are 
oblate spheroids with the short axis parallel to the direction
of the motion of the perturber.  

For supersonic motion ($\mach>1$), we seek for asymptotic solutions 
of $\D$ just behind the perturber along its orbit (i.e., 
$0<-s\ll 1$, $\tR=1, \tz=0$).
We further restrict the discussion to not-so-highly 
supersonic cases\footnote{More
precisely, $1<\mach  < \mach_1$, where $\mach_1$ is a critical Mach
number discussed in Appendix \ref{sec:wi}.}, 
for which equation (\ref{eq:w}) has three roots:
\begin{mathletters}\label{eq:wthree}
  \begin{eqnarray}
    w_1 &\approx& -\frac{s}{1 + \mach}, \\
    w_2 &\approx& -\frac{s}{1 - \mach}, \\
    w_3 &\approx& w_0 -s \left(1 - \frac{\mach^2 \sin w_0}{w_0}\right)^{-1},
  \end{eqnarray}
\end{mathletters}
as long as $0<-s \ll (\mach-1)^{3/2}$. 
Here, $w_0$ is a non-zero solution of
\begin{equation}
\mach(2-2\cos w_0)^{1/2} + w_0=0
\end{equation}
for given $\mach$ ($\neq1$). 
Note that $w_1$ and $w_2$ are analogous to two solutions of O99 that
contribute the wake in the rear Mach cone for a linear-trajectory
perturber, while $w_3$ is a new solution introduced by a circular orbit.
For $\mach \rightarrow 1^{+0}$, $w_0 \rightarrow -[24(\mach-1)]^{1/2}$.

Substituting equations (\ref{eq:wthree}) into equation (\ref{eq:D})
and taking the steady-state limit ($\step \rightarrow 1$),
we obtain
\begin{equation}\label{eq:Dss}
\D(s) = \frac{2}{\vert s \vert} + 
\frac{\mach}{\mach^2\sin w_0 - w_0}, 
\end{equation}
for $-s \ll (\mach-1)^{3/2}$.
The first term in the right-hand side of equation (\ref{eq:Dss}) again 
represents the perturbed density in the linear trajectory case. 
The second term, albeit small compared with the first term for small $|s|$, 
arises as the perturber on a circular orbit can overtake its wake.  
The effects of circular orbits are significant in regions
away from the perturber, which is pursued numerically in \S\ref{sec:res}.

\section{Number of Roots for $w_i$ of Equation (\ref{eq:w})} \label{sec:wi} 

To obtain the perturbed density $\D(\vct{x},t)$ at given time and location,
it is crucial to find $w_i$ that satisfies equation (\ref{eq:w}).
Let us define $f(w)\equiv \mach d(w; \tR, \tz)$ for a fixed set of 
$\mach, \tR, \tz$, and $s$.  Then, finding solutions of equation 
(\ref{eq:w}) is equivalent to finding the intersections of two 
curves $y = -w -s$ and $y=f(w)$ on the $w$ -- $y$ plane.  Since 
$\vert df/dw \vert \leq \mach$ for all $\tR$ and $\tz$, it is 
trivial to show that
there is only one solution for all space and time if $\mach < 1$.
This implies that any location inside the casual 
region for sound waves is influenced just once by a perturber moving 
at a subsonic speed.  The same is true for a 
subsonic perturber moving on a straight-line trajectory studied 
by O99.

A necessary condition for having more than one root is 
$\vert df/dw \vert > 1$ for any $w$, which after some algebra results in 
\begin{equation}\label{eq:ncond}
    \mach > 1, \quad
    \tR   > \mach^{-1},\quad\mbox{and} \quad
    |\tz|   < \mach^{-1}\ \sMsml\ \sMsRsml\mbox{.}
\end{equation}
When the conditions (\ref{eq:ncond}) are fulfilled, there are two
tangent points of the curve $y=f(w)$ to lines with 
a slope of $-1$ in the range $-\pi < w < 0$.  
Let ($w_+$, $y_+$) and ($w_-$, $y_-$) denote the tangent points, 
such that 
$-\pi < w_- < w_+<0$ and $0<y_+ < y_-<\mach$.  
The tangent condition $df/dw\vert_{w_\pm} =-1$ yields 
\begin{mathletters}\label{eq:wy}
  \begin{eqnarray}
    w_\pm &=& - \cos^{-1}
          \left(\frac{1\pm\sMsml\sMsRsml}{\mach^2\tR}\right)\mbox{,} \\
    y_\pm &=& \vert \sMsRsml \mp \sMsml \vert,
  \end{eqnarray}
\end{mathletters}
at the $z=0$ plane.
Note that the $2\pi$ periodicity of $y=f(w)$ ensures that 
($w_\pm + 2\pi n$, $y_\pm$) for arbitrary integer $n$ are also tangent 
points to a line with slope of $-1$.

For given $\mach>1$,
the number of solutions of equation (\ref{eq:w}) depends on
the slopes of lines connecting the points $(w_- - 2\pi n, y_-)$ and 
$(w_+, y_+)$.  When the line passing through $(w_- - 2\pi, y_-)$ and 
$(w_+, y_+)$ has a slope larger than $-1$, which occurs for
$1 < \mach < \mach_1$ (see below for the definition of $\mach_n$ with 
integer $n\geq 1$), the spatial regions that satisfy equations
(\ref{eq:ncond}) as well as 
$ w_+ + y_+ < -s < w_- + y_-$ have three roots for 
$w_i$, while the other regions have only a single root.
When $\mach_1 < \mach < \mach_2$, the slope of the line connecting 
$(w_- - 4\pi, y_-)$ and $(w_+, y_+)$ is greater than $-1$, so that the 
regions bounded by the condition $ w_+ + y_+ < -s < w_- + y_- - 2\pi$
possess five roots; the other regions 
have three roots.  Note that the volume that does not satisfy conditions
(\ref{eq:ncond}) still has a single root.
As the Mach number increases further, new regions emerge to have
a larger (odd) number of roots for $w_i$.  
Generalizing the discussion given above,
the critical Mach number $\mach_n$ below which a steady-state wake 
possesses up to $2n+1$ roots can be determined  by the requirement
$(y_+ - y_-)/(w_+ - w_- + 2\pi n) =-1$.  Using 
equations (\ref{eq:wy}), this is simplified to
\begin{equation} \label{eq:M}
  2 (\mach_n^2 -1)^{1/2}-\cos^{-1}\,(2\mach_n^{-2}-1) = 2 \pi n,
\end{equation}
where we take $\tR=1$ without any loss of generality since the existence
of multiple roots requires $\tR > \mach^{-1}$ (eq. [\ref {eq:ncond}]).
The first few solutions of equation (\ref{eq:M}) are $\mach_1=4.6033$,
$\mach_2=7.7897$, $\mach_3=10.9499$, etc.

The presence of multiple roots implies that the density wake in those 
regions are constructed by a superposition of sonic signals that were 
emitted by the perturber at as many different locations (and 
corresponding retarded times) 
as the number of roots.
The $\mach$-dependency of the number of roots for a circular-orbit
supersonic
perturber is in sharp contrast to the case of a straight-line trajectory
where only one (within a sonic sphere) or two (within
a rear Mach cone) roots are allowed regardless of the Mach number (O99).
When a perturber is moving 
at a supersonic speed along a circular orbit, it is able to catch up with
its own wake (possibly multiple times), adding new sonic disturbances
to the regions that were already disturbed by the perturber.
Regions of multiple roots 
usually take a form of a long trailing spiral, as exemplified in Figures 
\ref{evolution},  \ref{steadyM20}, and  \ref{overlap}.

\section{HALF-OPENING ANGLE} \label{sec:angle}

\begin{figure}
\epsscale{0.8}
  \plotone{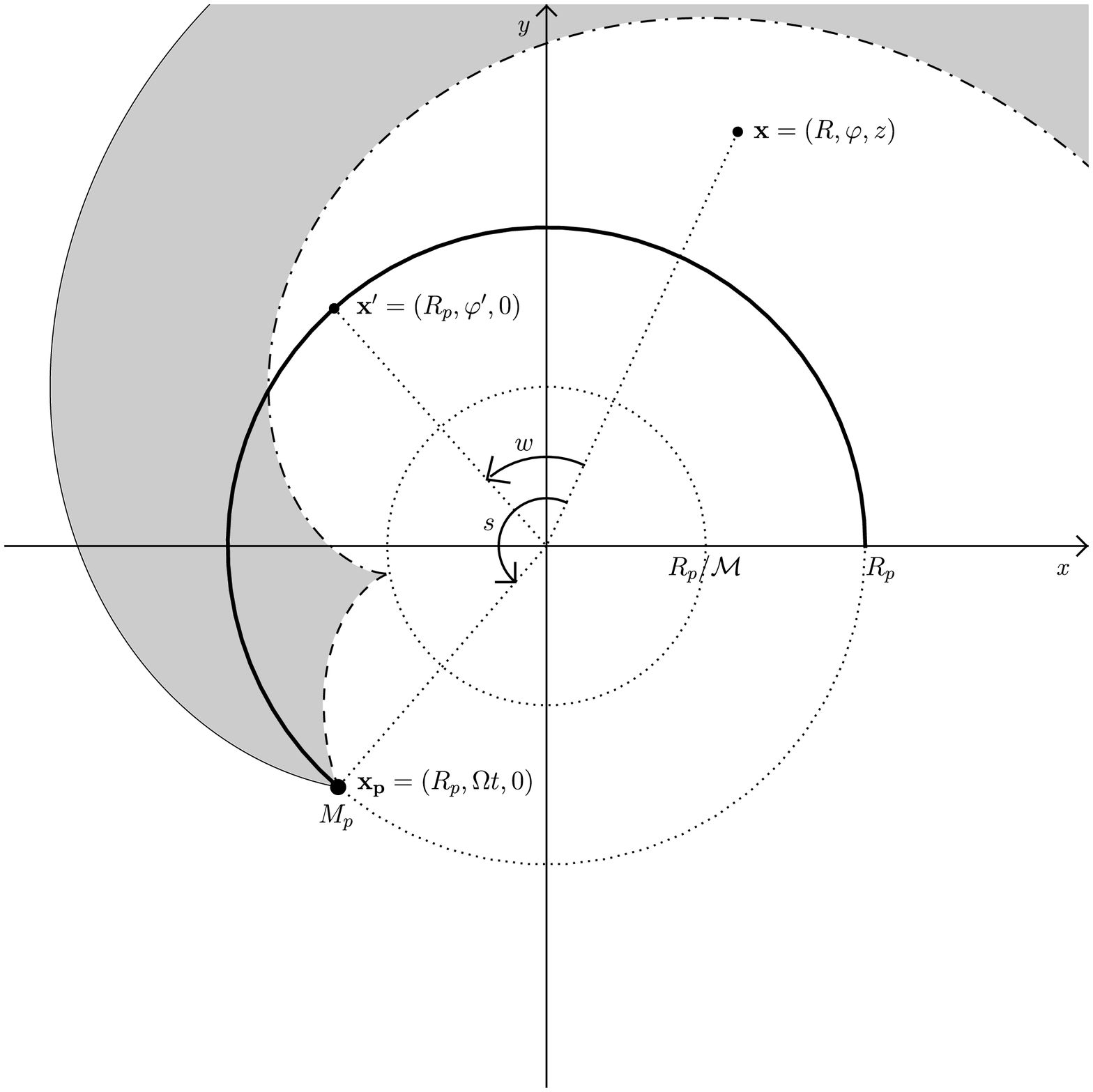}
  \caption{\label{sketch}
Schematic diagram illustrating the situation on the $z=0$ plane at 
time $t$.  A perturber initially introduced at $(R_p, 0, 0)$ moves 
along a circle with radius $R_p$ at uniform angular speed
$\Omega$ in the counterclockwise direction, and is currently at
the position $\vct{x}_p\equiv (R_p, \Omega t, 0)$.
At this time, an observer sitting at $\vct{x}=(R, \varphi, z)$ receives 
a sonic signal that was emitted by the perturber when it was at 
$\vct{x'}=(R_p, \Omega t', 0)$, where 
$t'=t-\vert\vct{x}-\vct{x'}\vert/\cs$ is the retarded time. 
The angular variables are $w\equiv \varphi'-\varphi$ and 
$s\equiv \varphi-\Omega t$ along the orbital plane. 
The shaded area represents a curved Mach cone and a wake tail 
formed by a supersonic perturber with $\mach>1$.  
The inner edge (dashed curve) of the Mach cone
meets the inner edge (dot-dashed curve) of the tail 
at a point on a circle with radius $R_p/\mach$.
The outer edge (light solid curve) of the Mach cone defines 
the outer edge of the tail.}
\end{figure}

\begin{figure}
\epsscale{1.0}
  \plotone{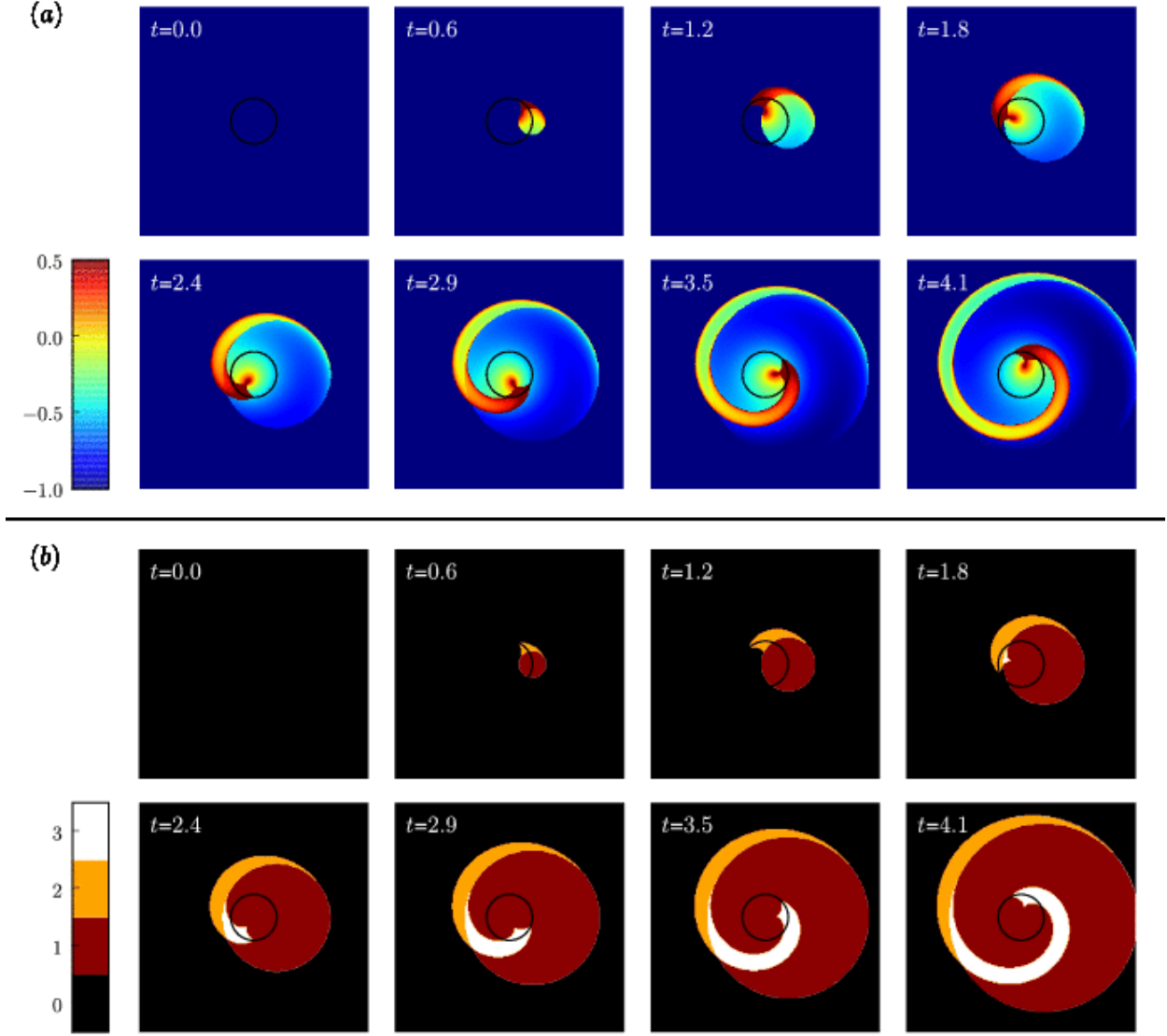}
  \caption{\label{evolution}
Temporal evolution at the $z=0$ plane of (a) the 
dimensionless density wake $\D(\vct{x},t)$
in logarithmic color scale and (b) 
the number of roots for $w_i$ in equation (\ref{eq:w}) 
for the case of $\mach=2.0$.
The time is in unit of $R_p/\cs$.  See text for details.}
\end{figure}

\begin{figure*}
  \plotone{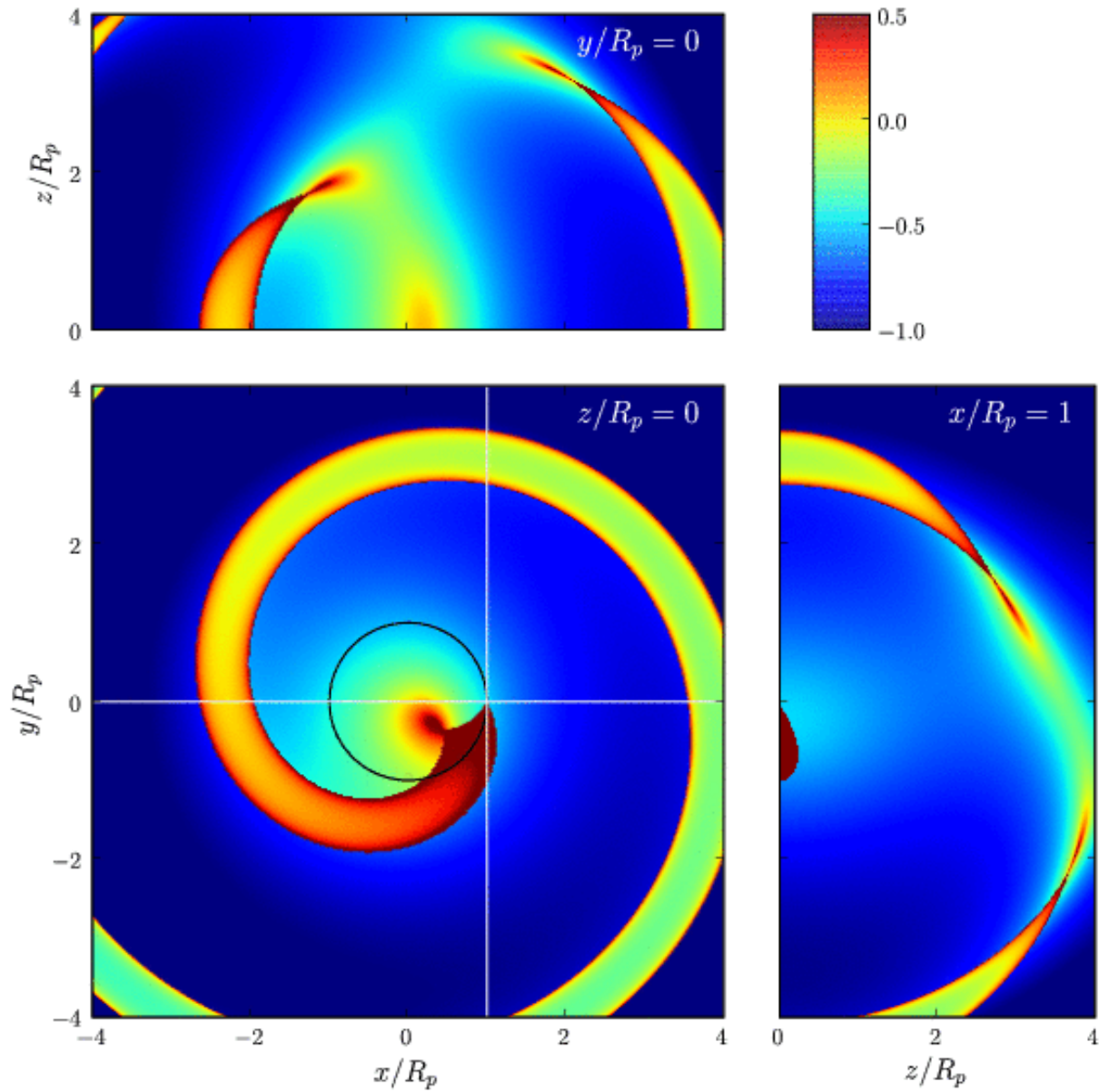}
  \caption{\label{steadyM20} 
Density distributions of the steady-state wake for $\mach=2.0$ 
on the $x/R_p=1$ (\textit{bottom right}), 
$y=0$ (\textit{top left}), and $z=0$ (\textit{bottom left}) planes.  
The perturber is located at $(x,\,y)=(R_p,\,0)$,
and the black circle in the bottom left frame denotes the orbit of
the perturber.  Color bar labels $\log \D$.}
\end{figure*}

\begin{figure*}
\epsscale{1.}
  \plotone{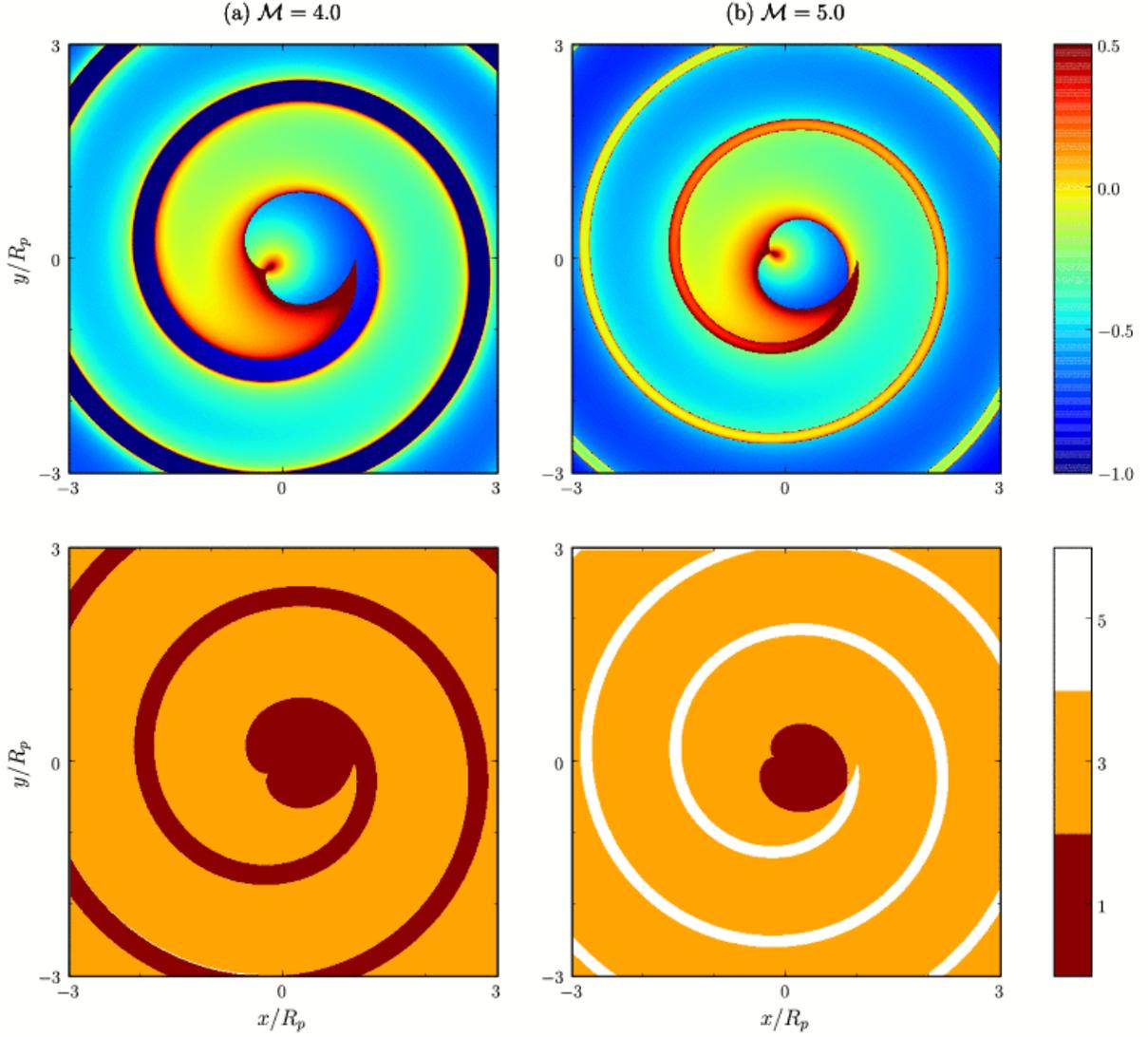}
  \caption{\label{overlap}
Distributions of the perturbed density $\D$ in logarithmic color scale 
(\textit{top}) and the corresponding number of roots for 
$w_i$ in equation (\ref{eq:w}) (\textit{bottom})
at the $z=0$ plane of the steady-state wake for 
(a) $\mach=4.0$ and (b) $\mach=5.0$.
The perturber is located at $(x,\,y)=(R_p,\,0)$.}
\end{figure*}

\begin{figure}
  \plotone{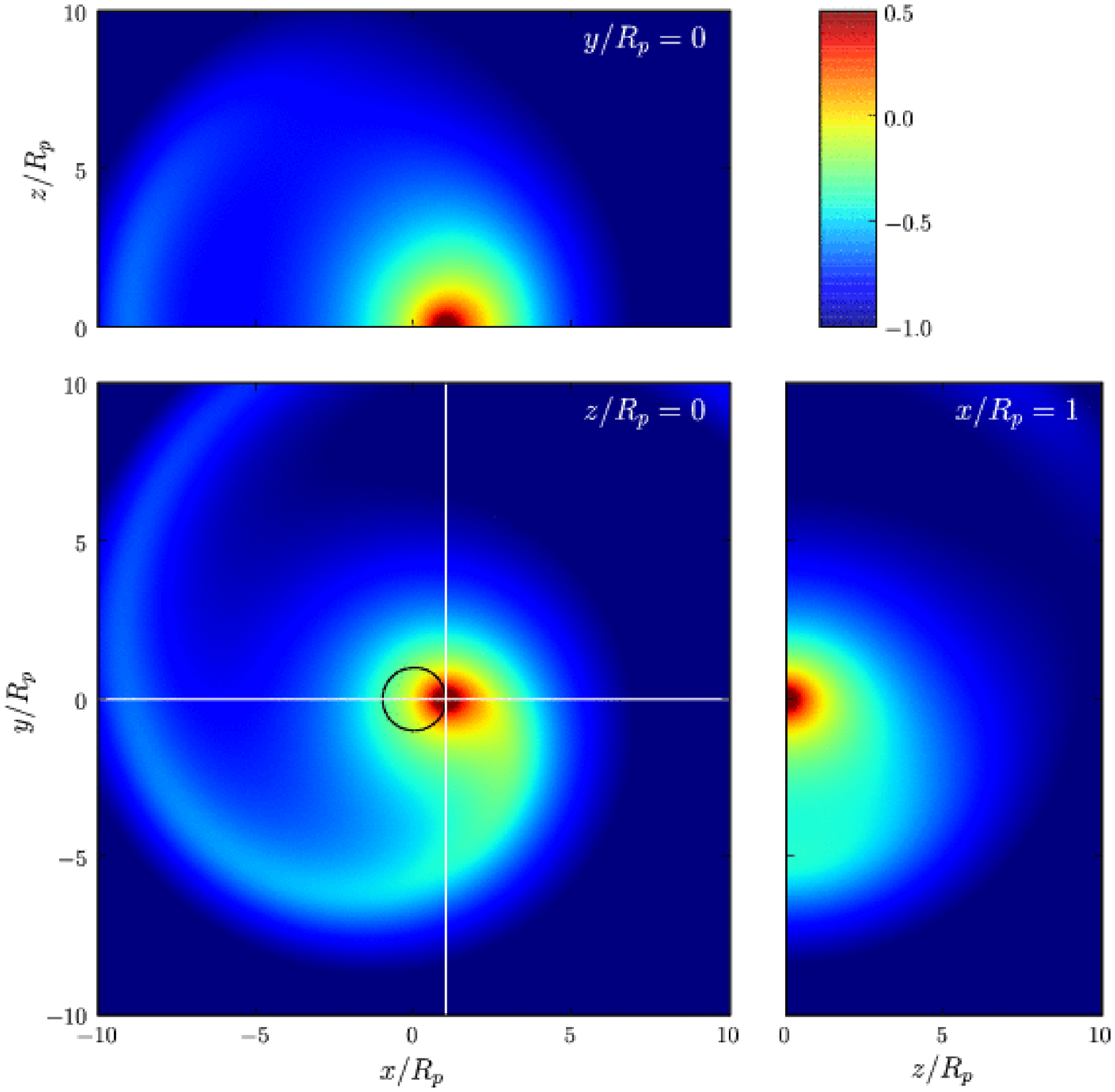}
  \caption{\label{steadyM05} 
Same as Fig.~\ref{steadyM20} except for $\mach=0.5$.}
\end{figure}

\begin{figure}
  \plotone{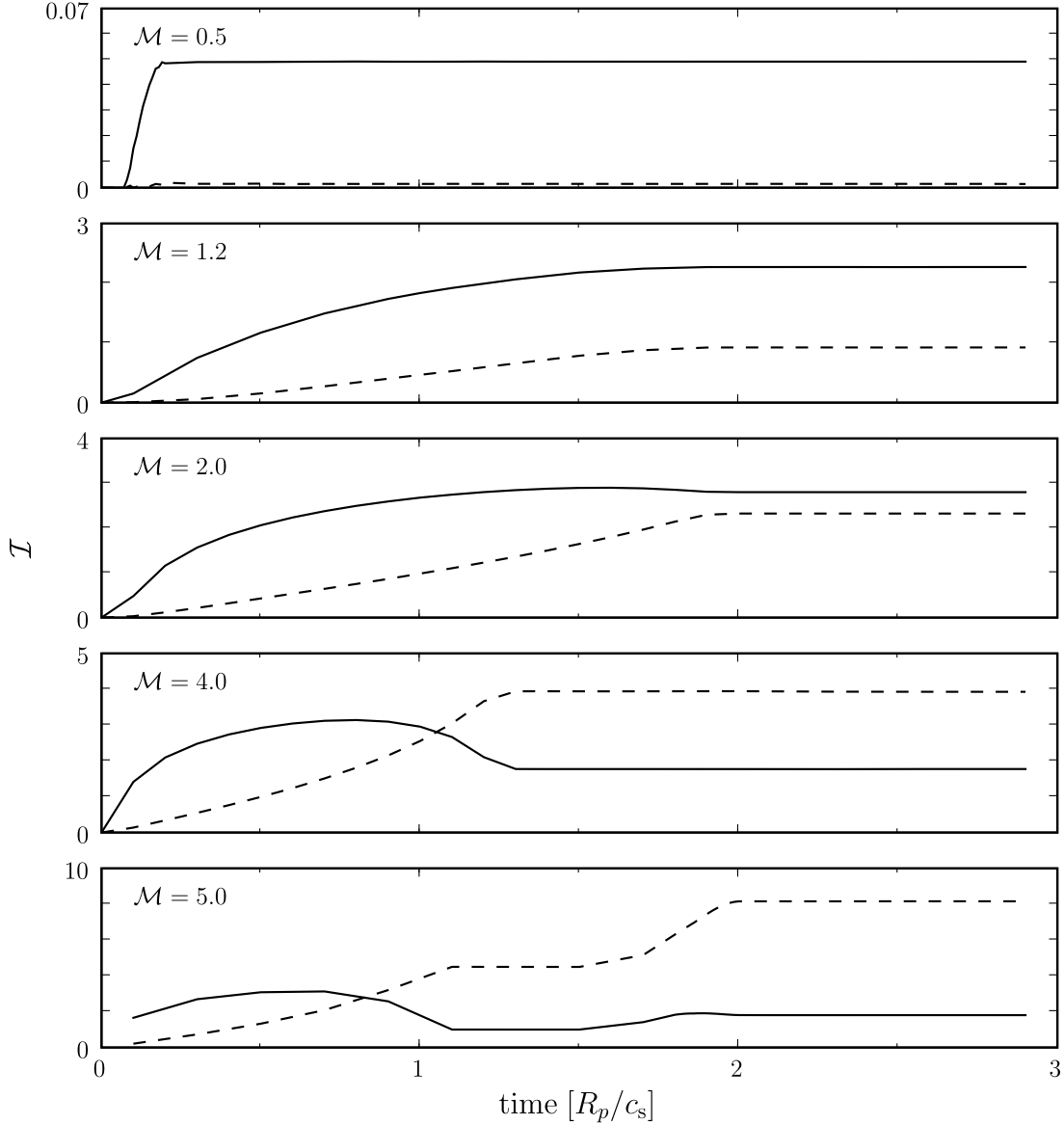}
  \caption{\label{time}
Time evolution of the drag force for $\mach=0.5$, 1.2, 2.0, 4.0, and 5.0. 
Solid curves draw the azimuthal drag $\I_\varphi$, while
dashed curves are for the radial drag $\I_R$. 
For all cases, $\rmin=R_p/10$ is taken.  Note that 
both $\I_\varphi$ and $\I_R$ converge typically within the
sound crossing time across $2R_p$, 
indicating that a steady state is reached quite rapidly.}
\end{figure}

\begin{figure} 
\epsscale{1.}
  \plotone{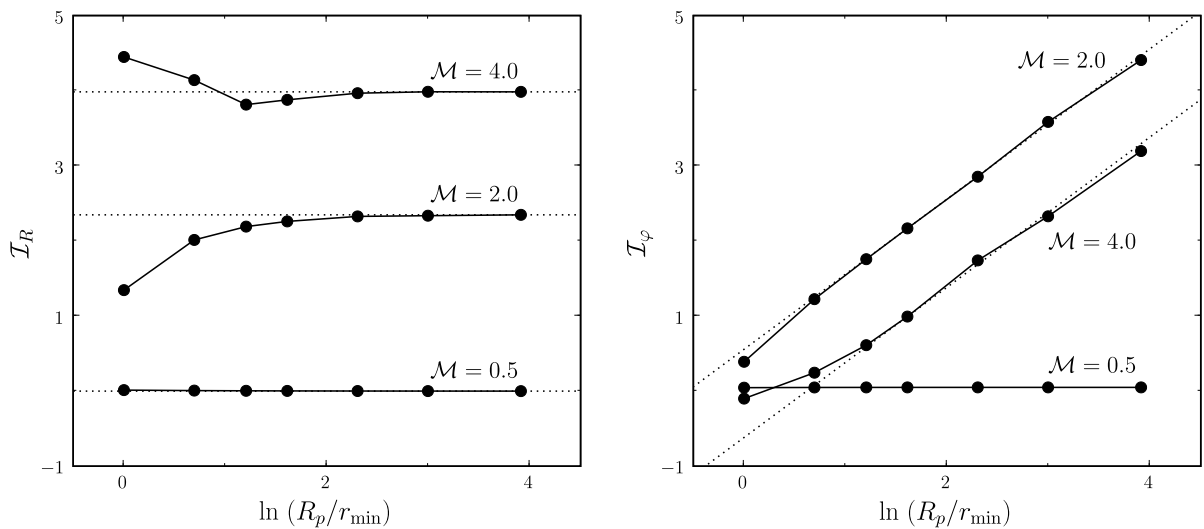}  
  \caption{\label{rmin}
Dependence on $\rmin$ of the steady-state drag force 
for $\mach=0.5$, 2.0, and 4.0.
\textit{Left}: For $R_p/\rmin > 10$, the radial drag force
$\I_R$ converges to a respective constant value marked by dotted line.
\textit{Right}: The azimuthal drag force $\I_\varphi$ 
varies as $\ln\,(R_p/\rmin)$  
for small $\rmin$ when $\mach>1$, while independent of $\rmin$ 
for $\mach<1$.  Dotted lines indicate a slope of unity.}
\end{figure}

\begin{figure} 
\epsscale{1.}
  \plotone{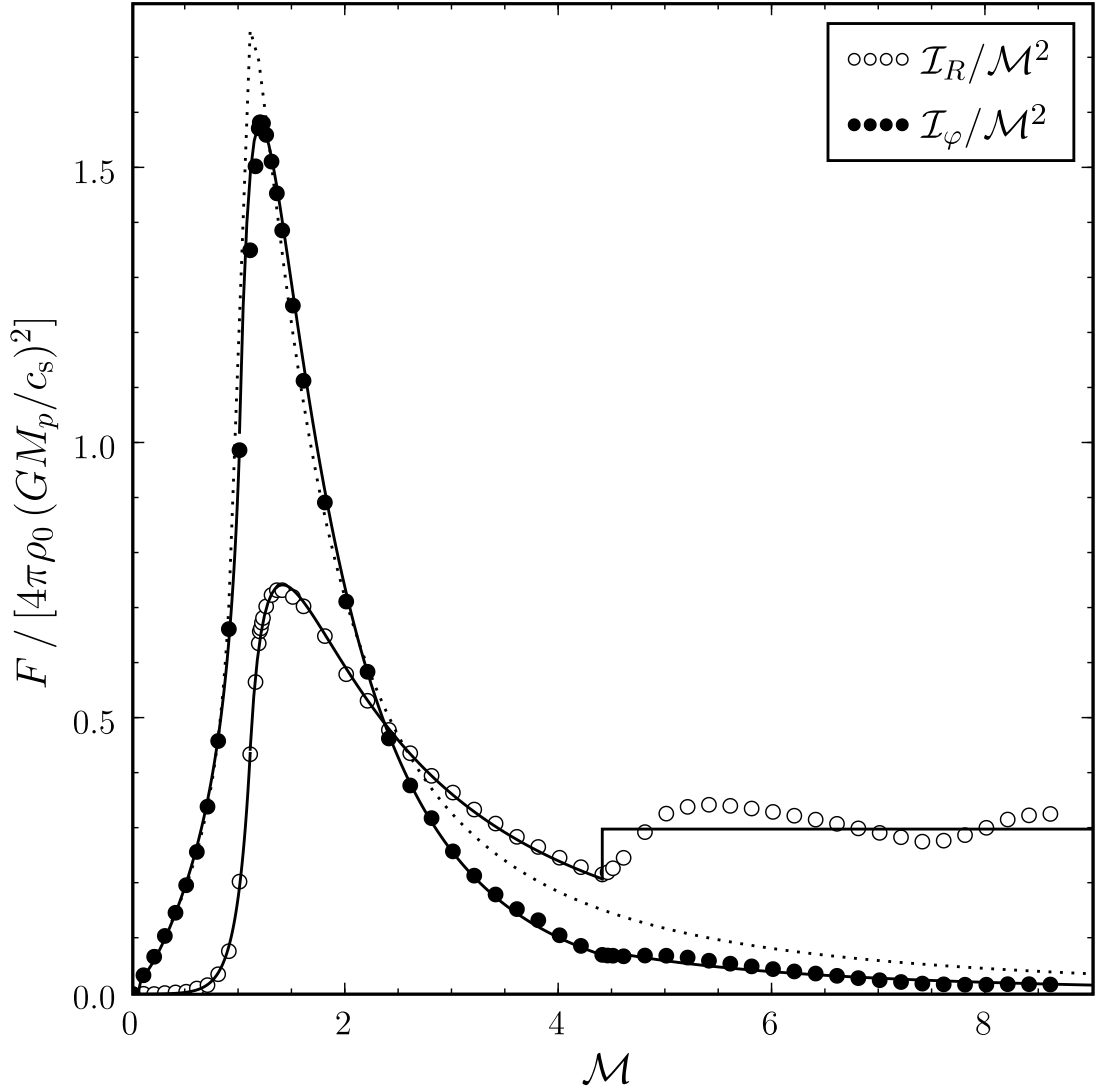}  
  \caption{\label{forces}
Gravitational drag force on a circular-orbit perturber in a gaseous
medium as a function of the Mach number $\mach$.  
The open and filled circles give the results of our semi-analytic 
calculation for the drag in the radial and azimuthal directions,
respectively.  
For all the points, $\rmin/R_p=0.1$ is taken.
Solid lines draw the fits, equations (\ref{eq:IR}) and (\ref{eq:Iphi}),
to the semi-analytic results.  Dotted line corresponding to 
the force formula with $V_p t=2R_p$ in O99 for the case of 
linear-trajectory perturbers is in quite a good agreement with the 
azimuthal drag for circular-orbit perturbers.}
\end{figure}

\begin{figure}
\epsscale{1.}
  \plotone{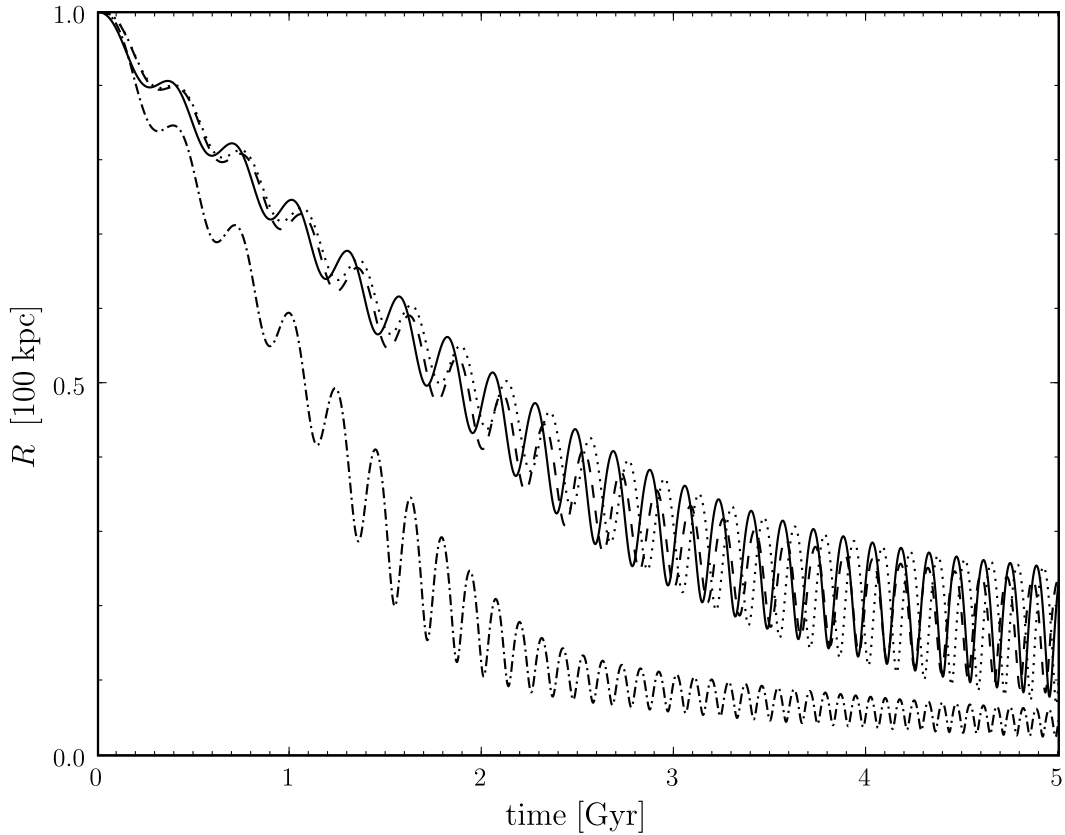}
  \caption{\label{orbitNFW}
Orbital decay of a model galaxy caused by dynamical friction due to an
intracluster gas. Solid line corresponds to the case when 
both $\I_R$ (eq.\ [\ref{eq:IR}]) and 
$\I_\varphi$ (eq.\ [\ref{eq:Iphi}]) are considered, while dashed line shows 
the result with only $\I_\varphi$ (i.e., with $\I_R=0$).  
For comparison, the results of Ostriker's formula with fixed $V_p t=1$ Mpc 
and varying $V_p t = 2R(t)$ are plotted as dot-dashed and dotted lines,
respectively.}
\end{figure}

Although a Mach cone is generally curved for a circular-orbit
perturber, its half-opening angle in the vicinity of the perturber 
is the same as in the linear trajectory counterpart.  
As explained in \S\ref{sec:superwake},  the interior of a curved Mach cone 
and an attached trailing tail is the region where equation (\ref{eq:w}) 
has multiple roots in a steady state.  It is bounded by 
$-s=w_\pm + y_\pm$ curves in the $z=0$ plane.  
More specifically, the $-s=w_+ + y_+$ curve describes the inner and outer 
edges of the
Mach cone head near the perturber as well as the outer edge of the tail, 
represented by light solid curve and dashed curve at the
boundaries of the shade region displayed in \Fig{sketch}.
On the other hand, the $-s=w_- + y_-$ curve corresponds to the inner 
edge of the tail (dot-dashed curve in Fig.~\ref{sketch}) that 
meets the Mach cone head at $\tR = \mach^{-1}$.
By Taylor expanding $w_+$ and $y_+$ about $\tR=1, s=0$ 
in the $z=0$ plane and adding the resulting expressions together, 
one can show the head of the Mach cone is described by
\begin{equation}\label{eq:half}
  s = - (\mach^2 -1)^{1/2} \vert \delta \vert,
\end{equation}
up to the first order in $\vert \delta \vert\equiv \vert \tR-1\vert$.  
The negative sign in equation (\ref{eq:half})
results since $s$ is measured in the counterclockwise direction 
from the perturber
(Fig.~\ref{sketch}).  Let $\theta$ be the half-opening angle of the Mach 
cone near the perturber. Then, $\tan \theta =  -|\delta|/s
= (\mach^2-1)^{-1/2}$, or $\theta =  \sin^{-1}(1/\mach)$, the same result
as in the linear trajectory case.

\clearpage

\clearpage

\clearpage

\clearpage

\clearpage

\clearpage

\clearpage

\clearpage

\clearpage

\end{document}